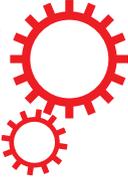

# SCIENTIFIC REPORTS

**OPEN**

# Electric control of superconducting transition through a spin-orbit coupled interface

Jabir Ali Ouassou[1], Angelo Di Bernardo[2], Jason W. A. Robinson[2] & Jacob Linder[1]

We demonstrate theoretically all-electric control of the superconducting transition temperature using a device comprised of a conventional superconductor, a ferromagnetic insulator, and semiconducting layers with intrinsic spin-orbit coupling. By using analytical calculations and numerical simulations, we show that the transition temperature of such a device can be controlled by electric gating which alters the ratio of Rashba to Dresselhaus spin-orbit coupling. The results offer a new pathway to control superconductivity in spintronic devices.

The dissipationless flow of electric charge and phase coherence are major driving forces for the research and development of superconducting electronics along with phase coherence. For example, superconducting logic circuits have already been implemented, including computer processors and memory chips that work at frequencies up to several gigahertz[1–4]. For spintronics[5–8] the main aim is to create logic and memory devices that exploit both the charge and spin degrees of freedom of electrons, and which offer high operating frequencies and low energy consumption[9].

In recent years, there has been a surge of interest in the intersection of these fields, and new discoveries have enabled the new field of superconducting spintronics[10,11]. At the interface between a conventional superconductor and a ferromagnet, the singlet electron pairs $|{\uparrow\downarrow}\rangle - |{\downarrow\uparrow}\rangle$ in the superconductor can be transformed into spin-polarized triplet pairs through a two-step process involving *spin-mixing* and *spin-rotation*[10]. Spin-mixing occurs at magnetic interfaces whereas magnetic inhomogeneities[12,13] or spin-orbit coupling[14–16] can rotate triplet Cooper pairs into each other, leading to long-ranged proximity effects in strong ferromagnets[17–31].

One important application of superconducting spintronics is to control the temperature $T_c$ at which a material becomes superconducting using *spin-valves*[32–44]. These systems consist of a superconductor proximity-coupled to two ferromagnetic layers. By changing the relative magnetization direction of two ferromagnets one can toggle superconductivity on and off. A key to achieving this effect lies in whether the magnetic configuration allows generation of spin-polarized Cooper pairs or not. When permitted, the generation of spin-polarized pairs which can penetrate deeper into adjacent ferromagnets opens an extra proximity "leakage channel". This contributes to the draining of superconductivity from the superconductor and therefore further reduces $T_c$. Although much research has been dedicated to magnetic control of $T_c$, it would be beneficial to be able to *electrically* control $T_c$, as that would enable integration of superconducting nanostructures into electronic circuits without the requirement of applying magnetic fields, *e.g.* by changing the quasiparticle distribution[45,46].

Here we propose a device comprised of a ferromagnetic insulator (FI) and a semiconductor with a two-dimensional electron gas (2DEG) in contact with a conventional superconductor (S). Experimentally, it is known that the Rashba and Dresselhaus spin-orbit coupling in a 2DEG can be tuned via a gate voltage[47–50]: this voltage can change the Rashba coefficient by a factor of 1.5–2.5 in thin-film structures based on GaAs or InAs[47–49], and up to a factor of ~6 in nanowires[50]. These results were obtained for different gate voltage ranges; e.g., ref. 48 varied it from −6 V to +2 V, while ref. 42 used −1.0 V to +1.5 V. It has also been shown that a suitably doped 2DEG can have Rashba and Dresselhaus coefficients of the same order of magnitude, with a ratio of ~1.5 in GaAs/AlGaAs[51]. It should therefore be possible to engineer a thin-film semiconductor with approximately matching Rashba and Dresselhaus couplings, and dynamically modulate the ratio between them by a factor of ~2 via a gate voltage.

[1]Department of Physics, NTNU, Norwegian University of Science and Technology, N-7491 Trondheim, Norway. [2]Department of Materials Science and Metallurgy, University of Cambridge, 27 Charles Babbage Road, Cambridge CB3 0FS, United Kingdom. Correspondence and requests for materials should be addressed to J.A.O. (email: jabir.a.ouassou@ntnu.no)

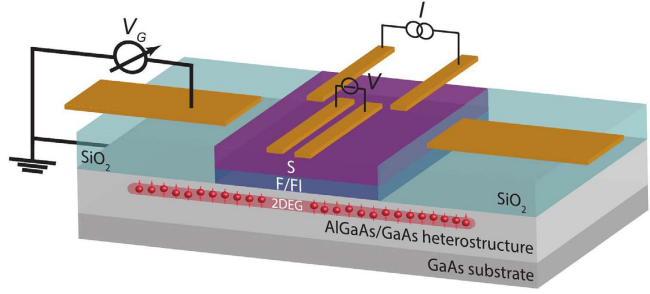

**Figure 1. Schematic of the proposed superconducting device.**

Recently, it was demonstrated that the superconductor proximity effect depends strongly on the amount of Rashba and Dresselhaus coupling that is present in the system[52]. Because of this, we set out to determine if $T_c$ could be controlled purely electrically by tuning the ratio of Rashba to Dresselhaus interactions with an electric field when a 2DEG is in electrical contact to a superconductor via a FI, the latter one serving as a source of triplet pairs. In this work, we confirm this conjecture and predict that all-electric control of $T_c$ is possible in a S/FI/2DEG device. In addition to a gate voltage control, $T_c$ also responds to a change in the FI magnetization orientation, causing our proposed device to function as a combined superconducting transistor and magnetic spin-valve. By *superconducting transistor*, we mean a device where a gate voltage is used to switch on and off superconductivity in the structure, thus controlling to what extent a supercurrent can flow through the superconductor. This type of functionality is of interest since it corresponds to an electrically controlled transition from finite to zero resistance.

Here we investigate a setup where magnetism and spin-orbit coupling are split into two distinct layers rather than coexisting in the same material[52], the former being experimentally more feasible to achieve. Furthermore, whereas previous works have modelled the superconductor/ferromagnet interface using spin-independent tunneling boundary conditions, we here use the recently derived boundary conditions for strongly spin-polarized interfaces[53]. This means that spin-dependent tunneling, phase-shifts, and depairing effects for arbitrarily strong polarization are included in our new model whereas this has not been possible previously in the literature.

## Results

**Proposed experimental setup.** Our proposed experimental setup is sketched in Fig. 1. The electrically controlled superconducting switch is based on an S/FI bilayer grown on an epitaxial GaAs-based (*e.g.* AlGaAs/GaAs) semiconductor thin-film multilayer. To enable electrical control over the Rasha spin-orbit interaction in the 2DEG, Au gate electrodes are fabricated by electron-beam lithography on a few-nanometer-thick insulating $SiO_2$ layer deposited after the growth and lithographic patterning of the S/FI stack. A four-point probe setup is used to measure changes in the superconducting critical temperature as a function of applied gate voltage $V_g$. Although the insulating $SiO_2$ layer should minimize possible modulations in the Curie temperature $T_c$ of the FI driven by the applied gate voltage $V_g$, which can alone have an effect on the superconducting proximity effect, control samples without the FI layer should also be fabricated to exclude this possibility. We also note that the Rasha spin-orbit coupling is independent on the polarity of $V_g$[50]. In contrast, $V_g$ usually has an opposite effect on $T_c$, meaning that a positive $V_g$ normally enhances $T_c$, while a negative $V_g$ decreases $T_c$[54]. Therefore, modulations in $T_c$ due to $V_g$ can also be excluded by investigating variations in the spin-orbit-driven superconducting proximity effect as a function of the $V_g$ polarity.

**Analytical results.** To explain the mechanism of the electric control of $T_c$, we first approximate the multilayer structure as an effective monolayer structure where spin-orbit coupling and magnetic exchange fields coexist. This analogy is relevant because the spin-dependent phase-shifts induced by proximity to a FI are known to act as an effective exchange field in thin superconducting structures[55]. Afterwards, we will confirm the analytical treatment by full numerical simulations performed without these approximations.

To the linear order in the superconducting pair amplitudes, the diffusion equations of the system are[52]

$$(iD/2)\partial_z^2 f_s = \epsilon f_s + h f_{\parallel}, \tag{1}$$

$$(iD/2)\partial_z^2 f_{\parallel} = E_{\parallel}(\theta,\ \chi) f_{\parallel} - R(\theta,\ \chi) f_{\perp} + h f_s, \tag{2}$$

$$(iD/2)\partial_z^2 f_{\perp} = E_{\perp}(\theta,\ \chi) f_{\perp} - R(\theta,\ \chi) f_{\parallel}. \tag{3}$$

The symbols $f_s$, $f_{\parallel}$, $f_{\perp}$ refer to the electron pair amplitudes with spin-singlet, short-range spin-triplet, and long-range spin-triplet projections, respectively. We have also defined the the triplet mixing factor

$$R(\theta,\ \chi) = 2iDA^2\ \cos 2\theta \sin 2\chi, \tag{4}$$

and the effective triplet energies

$$E_\perp(\theta,\ \chi) = \epsilon + 2iDA^2[1 + \ \sin 2\theta \ \sin 2\chi], \tag{5}$$

$$E_\parallel(\theta,\ \chi) = \epsilon + 2iDA^2[1 - \ \sin 2\theta \ \sin 2\chi]. \tag{6}$$

The ferromagnetism is described by an in-plane exchange splitting $\boldsymbol{h} = h(\cos\theta\,\boldsymbol{e}_x + \sin\theta\,\boldsymbol{e}_y)$, which is parametrized in terms of a magnitude $h$ and direction $\theta$. We also assume an in-plane spin-orbit coupling, which is described in polar coordinates by a magnitude $A \equiv \sqrt{\alpha^2 + \beta^2}$ and type $\chi \equiv \mathrm{atan}(\alpha/\beta)$, where $\alpha$ and $\beta$ are the Rashba and Dresselhaus coefficients. The spin-orbit coefficients are defined by the single-particle Hamiltonian $H = \frac{\alpha}{m^*}(p_y\sigma_x - p_x\sigma_y) + \frac{\beta}{m^*}(p_y\sigma_y - p_x\sigma_x)$, where $m^*$ is the effective mass, $\boldsymbol{p}$ the momentum, and $\boldsymbol{\sigma}$ the spin. Finally, $D$ is the diffusion coefficient of the material, and $\epsilon$ is the quasiparticle energy.

The singlet component $f_s$ is produced in all conventional superconductors. When these pairs leak into the adjoining ferromagnet, eqs (1–3) show that a magnetic exchange splitting $h$ induces a nonzero short-ranged triplet component $f_\parallel$ as is well-known[13]. When spin-orbit coupling is present ($A \neq 0$), with both Rashba and Dresselhaus contributions ($\sin 2\chi \neq 0$), one also generates the long-range[15] triplet component $f_\perp$ so long as the magnetization direction satisfies $\cos 2\theta \neq 0$. It is the latter observation which offers several ways to control the long-ranged triplet generation. Firstly, since the triplet mixing term is proportional to $\cos 2\theta$, we may enable this mechanism by letting $\theta \to 0$, or disable it by letting $\theta \to \pm\pi/4$. Secondly, since the same term is also proportional to $\sin 2\chi$, where we defined $\chi = \mathrm{atan}(\alpha/\beta)$, the mechanism is enhanced for $\alpha \cong \beta$, but suppressed when $\alpha \ll \beta$ or $\alpha \gg \beta$. Since the magnetization direction $\theta$ can be changed using an external magnetic field, and the Rashba coefficient $\alpha$ can be changed using an external electric field, this means that the triplet mixing can be in principle be controlled using either a magnetic field by itself[56–60], an electric field by itself, or a combination thereof.

It is important to note that the spin-orbit coupling not only introduces a coupling between the different types of spin-polarized Cooper pairs, but that it also has a depairing effect. This is seen by how $A$ modifies the diagonal terms in the equations above, resulting in an alteration of the effective energies in eqs (5) and (6) associated with the superconducting correlation functions $f$. Imaginary terms in the effective energy can be interpreted as a destabilization and suppression of the given correlations, so the spin-orbit coupling can suppress either $f_\parallel, f_\perp$, or both, depending on the parameters $\chi$ and $\theta$. It follows from eq. (5) that increasing the magnitude of $A$ and $\sin 2\chi$ increases this pair-breaking effect, meaning that the same spin-orbit coupling that maximizes the triplet mixing also maximizes the depairing. However, while the mixing term is proportional to $\cos 2\theta$, the depairing terms are proportional to $\sin 2\theta$. A key observation which enables the purely electric control over $T_c$ is that for a fixed magnetization orientation $\theta$, the depairing energy is controlled by the ratio of Rashba and Dresselhaus spin-orbit coupling $\chi = \mathrm{atan}(\alpha/\beta)$. This argument is of importance since we from the numerical simulations find that the dominant effect of the spin-orbit coupling on the critical temperature is not the long-range triplet generation, but rather the short-range triplet suppression. In fact, the most extreme results were obtained for $\theta = \pm\pi/4$, which are precisely the configurations where the linearized diffusion equations disallow triplet mixing.

### Numerical results.

We have calculated $T_c$ numerically and the results are shown in Fig. 2. S is taken as conventional (e.g. Nb), the FI (e.g. GdN, EuO) is treated as a polarized spin-active interface, and the semiconducting layer (e.g. GaAs, InAs) is treated as a normal metal with a Rashba–Dresselhaus spin-orbit coupling. We used the Ricatti-parametrization[61] including the case of spin-orbit coupling[52] together with general magnetic boundary conditions[53] valid for arbitrary polarization of the interface region. We provide a detailed exposition of the computation of the critical temperature in the Methods section.

For all structures, we assumed a thickness of $0.65\xi$ for the superconductor and $0.15\xi$ for the 2DEG, where $\xi$ is the zero-temperature coherence length of a bulk superconductor. Assuming $\xi = 30\,\mathrm{nm}$, this would imply a superconductor thickness of $\sim$20 nm and thickness of $\sim$4 nm for the spin-orbit coupled layer. As for the magnitude of the spin-orbit coupling, we normalized both $\alpha$ and $\beta$ to $\hbar^2/\xi$. If the effective quasiparticle mass $m^*$ is assumed equal to the bare electron mass, and we again set $\xi = 30\,\mathrm{nm}$, we find that $\alpha, \beta = 1$ in dimensionless units corresponds to a coupling $\alpha/m^*, \beta/m^* = 2.2 \times 10^{-12}\,\mathrm{eV\,m}$. The spin-active interface was taken to have an experimentally realistic spin-polarization of 50%, a tunneling conductance $G_T/G \in \{0.2, 0.3\}$, and a spin-mixing conductance $G_\varphi/G_T = 1.25$, where $G$ is the bulk normal-state conductance of both materials (taken as equal for simplicity). We have run extensive $T_c$ calculations for other parameter values as well (not shown here), where we find qualitatively the same behavior as in Fig. 2, but quantitatively less variation if either the tunneling conductance $G_T$ is reduced, the spin-mixing conductance $G_\varphi$ is reduced, or the spin-polarization is increased. In particular, depending on the quality of the contact between the 2DEG and the FI layer, the tunneling conductance could be very small compared to the normal-state conductance, $G_T \ll G$.

The results in Fig. 2 display the same basic dependence on the magnetic field direction: the critical temperature is maximal when $\theta \to -\pi/4$, and minimal when $\theta \to +\pi/4$. It is interesting to note how spin-valve functionality is obtained in the present structure with just one magnetic layer, tuning $T_c$ from a maximum to minimum upon 90 degrees rotation of the magnetization. The magnitude of this variation depends strongly on the parameters. For strong spin-orbit coupling and moderate interface conductance, we see a variation of nearly $0.6T_{cs}$ in Fig. 2, where $T_{cs}$ is the critical temperature of a bulk superconductor. This corresponds to 5.5 K for niobium; for comparison, the current experimental record for spin-valve effects is around 1 K[62]. Furthermore, in the region where $\theta > 0$, the critical temperature drops to zero, which means that such a device could in principle function as a spin-valve even at absolute zero. Increasing the interface polarization or weakening the spin-orbit coupling diminishes this effect.

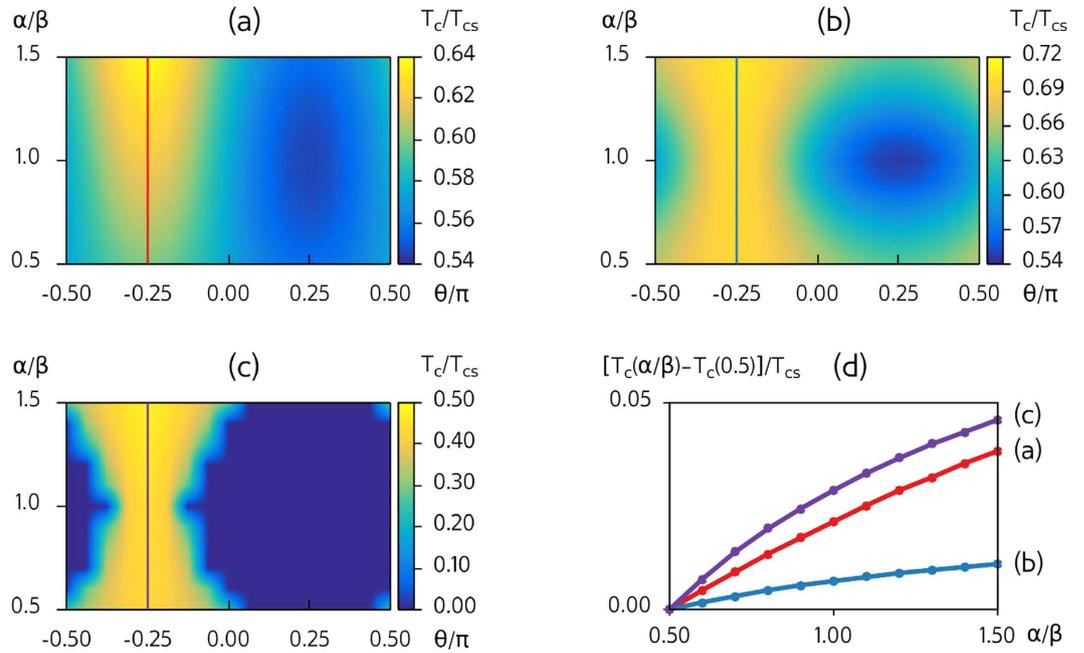

**Figure 2. Critical temperature results. (a–c)** Critical temperature normalized by the bulk value $T_c/T_{cs}$ (colors) as a function of the in-plane magnetization angle $\theta/\pi$ (horizontal axis) and spin-orbit ratio $\alpha/\beta$ (vertical axis). We have used the parameters **(a)** $\beta = 1$, $G_T/G = 0.2$, **(b)** $\beta = 5$, $G_T/G = 0.2$, and **(c)** $\beta = 5$, $G_T/G = 0.3$. **(d)** Variation $[T_c(\alpha/\beta) - T_c(0.5)]/T_{cs}$ in the critical temperature as a function of $\alpha/\beta$ when $\theta/\pi = -0.25$. The different curves correspond to the systems used in **(a–c)**. For other magnetization angles $|\theta/\pi| \neq 0.25$, the variation of $T_c$ with $\alpha/\beta$ is non-monotonic since such orientations allow for long-range triplet generation.

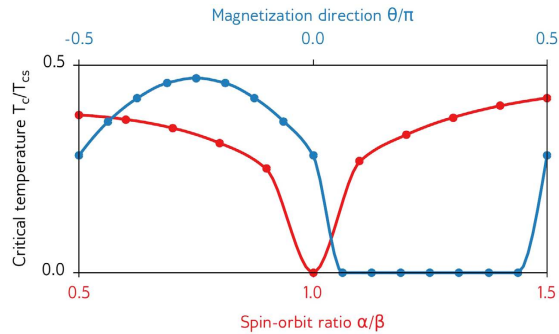

**Figure 3. Critical temperature highlights.** Normalized critical temperature $T_c/T_{cs}$ as function of the spin-orbit coupling ratio $\alpha/\beta$ (blue line, $\theta/\pi = -0.125$), and as function of the in-plane magnetization $\theta$ (red line, $\alpha/\beta = 1.5$). The other parameters are the same as in Fig. 2.

The most interesting observation is nevertheless that we can achieve all-electric control over $T_c$ for a fixed orientation $\theta$ of the FI magnetic moment. One particularly striking example is seen in Fig. 2: for a range of magnetization orientations $\theta$, $T_c$ increases from absolute zero at $\alpha/\beta = 1$ to a substantial fraction of the bulk critical temperature $T_{cs}$ as $\alpha/\beta$ is either increased or decreased. For instance, when $\theta/\pi = -0.125$, $T_c = 0$ at $\alpha/\beta = 1$ while $T_c = 0.42 T_{cs}$ at $\alpha/\beta = 1.5$. For e.g. niobium, this yields a variation of 3.9 K by increasing the Rashba coefficient $\alpha$ by 50%. We highlight this behavior in Fig. 3, where $T_c$ is plotted against the spin-orbit coupling ratio $\alpha/\beta$ for a fixed magnetization orientation. Moreover, we show in Fig. 3 the large change in $T_c$ that occurs when altering the in-plane magnetization orientation $\theta$ for a fixed $\alpha/\beta$.

Let us now interpret the numerical findings in terms of the previous analytical treatment. Although the 2DEG by itself has no intrinsic exchange field, rendering the distinction between short-ranged and long-ranged pairs more accurately described by the terminology "opposite and equal spin-pairing states relative the FI orientation", we will use to refer to $f_\parallel$ as short-ranged pairs for brevity and easy comparison with the analytical treatment. When $\alpha \to \beta$ and $\theta \to -\pi/4$, the short-ranged triplet energy $E_\parallel \to \epsilon + 4iDA^2$, resulting in a strong suppression of these triplet pairs. By closing the triplet proximity channel, this reduces the leakage of Cooper pairs from the superconductor, thus *increasing* the critical temperature of the structure. On the other hand, when $\alpha \to \beta$ and $\theta \to +\pi/4$, the energy $E_\parallel \to \epsilon$, resulting in a minimal suppression of short-ranged triplets. This causes a larger leakage from the superconductor, and *decreases* the critical temperature. This leading-order analysis of the physics is in accordance with the numerical results in Fig. 2, as is reasonable since the weak proximity effect described by the linearized equations is expected to be a good approximation for $T \cong T_c$.

## Discussion

In the quasiclassical theory used to compute the critical temperature, one assumes that the thickness of the layer exceeds the Fermi wavelength. This criterion is not satisfied in a 2DEG, which means that phenomena such as weak localization/antilocalization cannot be described by quasiclassical theory. However, the coupling mechanism governing the appearance of a superconducting triplet proximity channel in the system is not expected to change because of this and hence our results should remain qualitatively valid even in this scenario. Moreover, we have considered the diffusive limit of transport which is of relevance for the in-plane motion, whereas the 2DEG thickness is much smaller than the mean free path. There is nevertheless scattering at the multiple interfaces of our structure which is expected to enhance the effective diffusive character of quasiparticle motion considered in our model. 2DEGs can also feature a rather strong spin-orbit interaction, in which case corrections to the Usadel equation have been examined[63]. It could also be of interest to go beyond quasiclassical theory to study $T_c$ and other proximity effects in this kind of system[64], although this is beyond the scope of the present work.

Semiconductors such as GaAs and InAs are known to provide both an intrinsic Dresselhaus coupling and an electrically tunable Rashba coupling[48–50]. By combining such 2DEG materials with a superconductor and a ferromagnetic insulator, we have shown both analytically and numerically that $T_c$ responds to changes in both electric and magnetic fields, either individually or combined. It should therefore be possible to create a device that can function as a superconducting transistor, superconducting spin-valve, or both, depending on whether electric or magnetic stimuli are used as the input signal.

## Methods

**Diffusion equation.**    In the diffusive and quasiclassical limit, we can describe the structures discussed herein with the Usadel diffusion equation[15,16,52]

$$iD\,\check{\nabla}(\hat{g}\,\check{\nabla}\hat{g}) = [\epsilon\hat{\tau}_z + \hat{\Delta} + \hat{h},\ \hat{g}], \tag{7}$$

where $\hat{g}$ is the retarded quasiclassical propagator in Nambu $\otimes$ Spin space, $\hat{\tau}_z$ is the third Pauli matrix in Nambu space, $\epsilon$ is the quasiparticle energy of the electrons and holes, $\hat{\Delta} =$ antidiag $(+\Delta, -\Delta, +\Delta^*, -\Delta^*)$, $\Delta$ is the superconducting gap, $\hat{h} = \mathbf{h} \cdot \text{diag}(\boldsymbol{\sigma}, \boldsymbol{\sigma}^*)$, $\mathbf{h}$ is the ferromagnetic exchange field, $\boldsymbol{\sigma}$ is the Pauli vector in spin space, and $D$ is the diffusion coefficient. The notation $\check{\nabla}(\cdot) = \nabla(\cdot) - i[\check{A}, \cdot]$ is used for the gauge covariant derivative, where $\check{A} = \text{diag}(A, -A^*)$ is a background field that accounts for spin-orbit coupling. In this paper, we assume that we have a thin-film structure oriented along the $z$-axis so $\nabla \rightarrow \partial_z$. We assume the exchange field and Rashba–Dresselhaus coupling are both confined to the $xy$-plane, so they can be parametrized as

$$\mathbf{h} = h(\cos\theta\ \mathbf{e}_x + \sin\theta\ \mathbf{e}_y), \tag{8}$$

$$A = (\beta\sigma_x - \alpha\sigma_y)\mathbf{e}_x + (\alpha\sigma_x - \beta\sigma_y)\mathbf{e}_y. \tag{9}$$

Note that it is the orientation of the spin-orbit field $A$ that defines the $x$- and $y$-axes of our coordinate system, since the Dresselhaus spin-orbit coupling is determined by the crystal structure. Thus, the magnetic orientation is measured relative to the crystal structure.

Numerically, solving directly for the propagator $\hat{g}$ is impractical for two reasons. Firstly, the elements of $\hat{g}$ are unbounded, and can be arbitrarily large complex numbers. Secondly, the propagator satisfies a normalization condition and particle-hole symmetry which reduces the number of degrees of freedoms, such that solving for each individual matrix element in $\hat{g}$ would be redundant. Because of this, we have used the so-called Riccati parametrization of the propagators in the numerical simulations[61]:

$$\hat{g} = \begin{pmatrix} N & 0 \\ 0 & -\tilde{N} \end{pmatrix} \begin{pmatrix} 1 + \gamma\tilde{\gamma} & 2N\gamma \\ 2\tilde{N}\tilde{\gamma} & 1 + \tilde{\gamma}\gamma \end{pmatrix}, \tag{10}$$

where the Riccati parameters $\gamma$ and $\tilde{\gamma}$ are $2 \times 2$ matrices in spin space which are related by tilde-conjugation $\tilde{\gamma}(+\varepsilon) = \gamma^*(-\varepsilon)$, and the normalization matrices are defined by $N \equiv (1 - \gamma\tilde{\gamma})^{-1}$. For an in-plane spin-orbit interaction, eq. (7) parametrizes as

$$\begin{aligned} iD[\partial_z^2\gamma + 2(\partial_z\gamma)\tilde{N}\tilde{\gamma}(\partial_z\gamma)] &= 2\epsilon\gamma - \Delta\sigma_y + \gamma\Delta^*\sigma_y\gamma + \mathbf{h}\cdot(\boldsymbol{\sigma}\gamma - \gamma\boldsymbol{\sigma}^*) \\ &\quad + iD[AA\gamma - \gamma A^*A^* + 2(A\gamma + \gamma A^*)\tilde{N}(A^* + \tilde{\gamma}A\gamma)], \end{aligned} \tag{11}$$

which is the form we use numerically. For more information about the derivation and interpretation of the above, see ref. 52.

**Gap equation.**    Before we can calculate the critical temperature of a material, we require not only a way to calculate the propagator $\hat{g}$, but also a way to dynamically update the superconducting gap $\Delta$ based on the calculated propagators. This gap equation can be written[52]

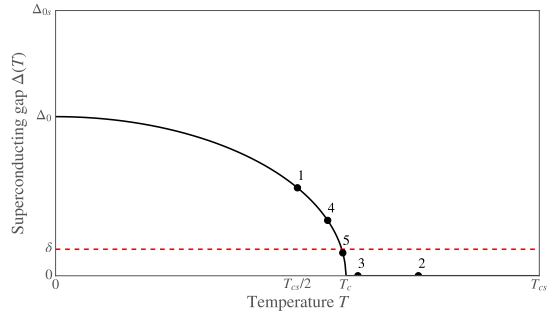

**Figure 4. Sketch of the superconducting gap $\Delta$ as a function of temperature $T$ for a superconducting hybrid structure.** When performing a binary search for the critical temperature, we check whether $\Delta > \delta$ at a certain number of temperatures–in other words, whether the black solid line is above the red dashed line. The numbered markers show which points on the curve would be evaluated during the first five bisections of such a binary search, and in what order. Note that since the algorithm is actually looking for the intersection between the black solid curve and red dashed curve, we need $\delta \ll \Delta_{0s}$ to obtain accurate results.

$$\Delta(z) = \lambda \int_0^{\Delta_{0s} \cosh(1/\lambda)} \mathrm{d}\epsilon \ \mathrm{Re}[f_s(\epsilon, \ z)] \tanh\left(\frac{\pi}{2e^c} \frac{\epsilon/\Delta_{0s}}{T/T_{cs}}\right), \tag{12}$$

where $\lambda$ is a dimensionless coupling constant, $f_s$ the singlet component of the anomalous propagator, $\Delta_{0s}$ the zero-temperature gap of a bulk superconductor, $T_{cs}$ the critical temperature of a bulk superconductor, and $c$ the Euler–Mascheroni constant. In terms of the Riccati parametrization, $f_s = [N\gamma]_{12} - [N\gamma]_{21}$, where the subscript notation refers to individual matrix elements.

**Boundary conditions.** Since our purpose is to model a system where the superconductivity, ferromagnetism, and spin-orbit coupling originate from *different* thin-film layers, we need boundary conditions that connect the propagators of these materials at the interfaces. Numerically, we focused on S/FI/N structures, in which case the ferromagnetic insulator itself is modelled as a strongly polarized spin-active interface. We have used the low-transparency limit of the general spin-active boundary conditions derived in ref. 53,

$$2\hat{I}_\mathrm{L} = G_\mathrm{T}[\hat{g}_\mathrm{L}, \ \hat{g}_\mathrm{R}] + G_1[\hat{g}_\mathrm{L}, \ \hat{m}\hat{g}_\mathrm{R}\hat{m}] + G_\mathrm{MR}[\hat{g}_\mathrm{L}, \ \{\hat{g}_\mathrm{R}, \ \hat{m}\}] - iG_\varphi[\hat{g}_\mathrm{L}, \ \hat{m}], \tag{13}$$

where $\hat{I}_\mathrm{L}$ is the $4 \times 4$ matrix current on the left side of the interface, $G_\mathrm{T}$ is the tunneling conductance of the interface, $G_1$ describes the interfacial depairing, $G_\mathrm{MR}$ describes the magnetoresistance, $G_\varphi$ describes the spin-mixing, $\hat{m} = \boldsymbol{m} \cdot \mathrm{diag}(\boldsymbol{\sigma}, \ \boldsymbol{\sigma}^*)$, $\boldsymbol{m}$ is a unit vector that describes the interface magnetization, and $\hat{g}_L$ and $\hat{g}_R$ describes the propagators on the left and right side of the interface, respectively. An equivalent equation for the other side of the interface can be found by letting $\hat{I}_\mathrm{L} \to -\hat{I}_\mathrm{R}$ and $L \leftrightarrow R$ in the equation above. Assuming that all the interface scattering have the same polarization $P$, it can be shown that $G_\mathrm{MR}/G_\mathrm{T} = P/\left[1 + \sqrt{1 - P^2}\right]$ and $G_1/G_\mathrm{T} = \left[1 - \sqrt{1 - P^2}\right]/\left[/1 + \sqrt{1 - P^2}\right]$, so we can calculate $G_1$ and $G_\mathrm{MR}$ directly from the interface polarization.

The matrix current is related to the propagators at the interface by $\hat{I} = GL(\hat{g} \tilde{\nabla} \hat{g})$ where $G$ is the normal-state conductance and $L$ the length of the material. It can then be shown that the Riccati parameters must satisfy the boundary condition

$$\partial_z \gamma = (2GLN)^{-1}(I_{12} - I_{11}\gamma), \tag{14}$$

where $I_{12}$ and $I_{11}$ refers to the top-right and top-left $2 \times 2$ blocks of the $4 \times 4$ matrix current $\hat{I}$. In this equation, the matrix current $\hat{I}$ should be interpreted as either $\hat{I}_\mathrm{L}$ or $\hat{I}_\mathrm{R}$, depending on which side of the interface the boundary conditions should describe.

**Critical temperature.** The critical temperature can be defined as the temperature $T_c$ such that the superconducting gap $\Delta = 0$ if and only if $T \geq T_c$. However, in practice, we cannot expect to obtain the exact result $\Delta = 0$ in simulations due to inexact numerical methods and random floating-point errors. For numerical simulations, we therefore use a more relaxed criterion $|\Delta| < \delta$ to define the critical temperature $T_c$, where we have set $\delta = 10^{-5}\Delta_{0s}$, and $\Delta_{0s}$ is the zero-temperature gap of a bulk superconductor. See the solid black curve in Fig. 4 for a sketch of how $\Delta(T)$ typically behaves, and how this is related to the critical temperature $T_c$.

Conceptually, the simplest way to find this critical temperature is to explicitly calculate the superconducting gap $\Delta$ as a function of temperature $T$, and check directly at which temperature we first find $|\Delta| < \delta$. However, such a linear search can be very costly when a high accuracy is desired. For instance, to determine the critical temperature to a precision of $0.0001T_{cs}$, where $T_{cs}$ is the critical temperature of a bulk superconductor, this would require that $\Delta(T)$ be calculated for 10,000 different values of $T$. For each of these temperatures, we need to solve a set of nonlinear diffusion equations for 150 positions and 800 energies, and repeat this procedure in one material of the hybrid structure at a time until a selfconsistent solution is found. Thus, the calculation at each of these 10,000 temperatures can in some cases take hours, making this method quite inefficient.

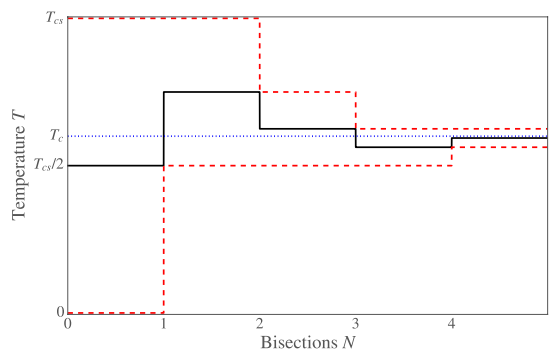

**Figure 5. Sketch of how the binary search algorithm works.** The blue dotted line at $T = T_c$ shows the critical temperature we wish to find, the solid black line shows the critical temperature estimate after $N$ bisections, and the red dashed lines shows the bounds for the critical temperature after $N$ bisections. After $N$ bisections, $T_c$ has been determined with an accuracy $T_{cs}/2^{N+1}$.

We have instead used a much more efficient binary search algorithm to determine the critical temperature numerically. The main benefit of this algorithm is that after calculating $\Delta(T)$ for $N$ particular values of $T$, we can determine the critical temperature to a precision $T_{cs}/2^{N+1}$. So in contrast to the linear search algorithm, an accuracy of around $0.0001 T_{cs}$ would require calculations at 12 temperatures instead of 10,000. Furthermore, we do not actually need to calculate $\Delta(T)$ exactly at these temperatures–it is sufficient to check whether $|\Delta| < \delta$ or $|\Delta| > \delta$ to determine whether $T > T_c$ or $T < T_c$. Thus, at each of these 12 temperatures, we only have to initialize the entire system to a BCS superconducting state with $\Delta = \delta$, then solve the Usadel equation and gap equation a fixed number of times in each material, and finally check whether $|\Delta| < \delta$ or $|\Delta| > \delta$ to determine whether $T$ is an upper or lower bound on $T_c$. How the binary search algorithm converges is illustrated in Figs 4 and 5.

## Acknowledgements

We wish to thank Niladri Banerjee for useful discussions. J.L. and J.A.O. acknowledge funding via the "Outstanding Academic Fellows" programme at NTNU, the COST Action MP-1201 and the Research Council of Norway Grant numbers 205591, 216700, and 240806. J.W.A.R. and A.D.B. acknowledge funding from the Leverhulme Trust (IN-2013-033), the Royal Society and the EPSRC through the Programme Grant "Superconducting Spintronics" (EP/N017242/1), and the Doctoral Training Grant (NanoDTC EP/G037221/1).

## Author Contributions

J.A.O. and A.D.B. conceived the idea, and J.A.O. performed the analytical and numerical calculations with support from J.L., J.A.O., A.D.B., J.W.A.R. and J.L. contributed to the discussion and writing of the manuscript.

## Additional Information

**Competing financial interests:** The authors declare no competing financial interests.

**How to cite this article**: Ouassou, J. A. *et al.* Electric control of superconducting transition through a spin-orbit coupled interface. *Sci. Rep.* **6**, 29312; doi: 10.1038/srep29312 (2016).